\documentclass[aps,twocolumn]{revtex4}
\usepackage[T1]{fontenc}
\usepackage[latin9]{inputenc}
\usepackage{amsthm}
\usepackage{amsmath}
\usepackage{graphicx}
\usepackage{multirow}
\usepackage{mathtools}
\usepackage{amssymb}
\usepackage{physics}
\usepackage{siunitx}

\begin{document}

\title{Effects of saturation and fluctuating hotspots for flow observables \\ in ultrarelativistic heavy-ion collisions}

\author{Henry Hirvonen, Mikko Kuha, Jussi Auvinen, Kari J. Eskola, Yuuka Kanakubo, Harri Niemi}

\affiliation{University of Jyv\"askyl\"a, Department of Physics, P.O. Box 35, FI-40014 University of Jyv\"askyl\"a, Finland,}
\affiliation{Helsinki Institute of Physics, P.O.Box 64, FI-00014 University of Helsinki, Finland}

\begin{abstract}
We investigate the effects of saturation dynamics on midrapidity flow observables by adding fluctuating hotspots into the novel Monte Carlo EKRT (MC-EKRT) event generator for high-energy nuclear collisions. We demonstrate that the intensity of the saturation effects significantly affects the ratio between the flow coefficients $v_3$ and $v_2$ at the LHC. Adding a hotspot substructure to the nucleons enhances the saturation effects and improves the agreement with the measured data. We show that the collision-energy dependence of the flow coefficients obtained using the MC-EKRT initial states with hotspots is improved in comparison with the earlier event-by-event EKRT model. In addition, we present the results for the charged hadron multiplicity distribution in Pb+Pb collisions at the LHC, and show that the minijet-multiplicity originating fluctuations of the saturation scale included in MC-EKRT, as well as the presence of hotspots, are necessary for describing the measured large-multiplicity tail in the distribution.
\end{abstract} 


\maketitle

\section{Introduction}

The highest-energy nucleus-nucleus collisions, ultrarelativistic heavy-ion collisions, which are currently performed at the CERN Large Hadron Collider (LHC) and at the Brookhaven National Laboratory (BNL) Relativistic Heavy Ion Collider (RHIC), aim at determining the properties of the nearly net-baryon-free hot Quark-Gluon Plasma (QGP). One also strives for a detailed understanding of the strong-interaction dynamics that is responsible for the creation and further evolution of the QGP in these collisions. See e.g.\ Ref.~\cite{ALICE:2022wpn} for a review.

According to lattice simulations of Quantum Chromodynamics (QCD, the theory of the strong interaction) the strongly-interacting matter takes  the form of the QGP at high temperatures of $T\gtrsim 150-160$~MeV \cite{Bazavov:2017dsy, Borsanyi:2013bia} at a vanishing baryochemical potential. Quarks and gluons can be produced in ultrarelativistic heavy-ion collisions from the kinetic energy of the colliding nuclei so copiously that the effective temperature (energy over particle ratio) of the system clearly exceeds 160 MeV. In these conditions, the normal formation of the color-confined, color-singlet bound states, hadrons, is momentarily inhibited, and a nearly-thermalized QGP, where the degrees of freedom are colored gluons, quarks and antiquarks, can be formed. The subsequent spacetime evolution stages of such a QCD matter -- the expansion and cooling of the QGP, the cross-over transition to a hadron gas, followed by the expansion and cooling of the hadron gas -- as well as the simultaneous appearance of the QGP and hadron-gas phases in different density regions of the expanding system, are describable in terms of relativistic dissipative fluid dynamics \cite{Romatschke:2007mq, Luzum:2008cw, Bozek:2009dw, Bozek:2012qs, Schenke:2010rr, Schenke:2010nt, Schenke:2011bn, Song:2011qa, Niemi:2011ix, Pang:2012he, Gale:2012rq, Niemi:2012ry, Noronha-Hostler:2013gga, Niemi:2015qia, Denicol:2015nhu, Ryu:2015vwa, Pang:2015zrq, Karpenko:2015xea, Giacalone:2017dud, Bozek:2017qir, Pang:2018zzo, Sakai:2020pjw, Nijs:2020roc, Shen:2022oyg}. While QCD is a cornerstone of the Standard Model of particle physics, relativistic fluid dynamics has become a standard tool in the analysis of heavy-ion observables.

The determination of the QCD matter properties, such as its equation of state and transport properties like the shear and bulk viscosities, from the measured LHC and RHIC observables is a highly challenging task. Clearly, a precise determination requires a simultaneous analysis of as many heavy-ion observables as possible, from as many collision systems and collision energies as possible -- a  ``global analysis'' of heavy-ion observables \cite{Niemi:2015qia, Gale:2012rq, Song:2011qa, Hirvonen:2022xfv}. A proper statistical analysis, Bayesian inference \cite{Novak:2013bqa, Bernhard:2016tnd, Bass:2017zyn, Bernhard:2019bmu, JETSCAPE:2020mzn, Nijs:2020roc, Auvinen:2020mpc, Parkkila:2021tqq, Parkkila:2021yha, Nijs:2023yab,Auvinen:2017fjw, JETSCAPE:2023nuf, Soeder:2023vdn} is necessary for setting well-defined uncertainties to the extracted matter properties. Interestingly, neural networks are currently making it possible to include also statistics-expensive observables, such as complicated rare flow correlators, into the global analysis \cite{Hirvonen:2023lqy, Hirvonen:2024ycx} (see also Ref.~\cite{Huang:2018fzn}).

The mentioned global analyses of heavy-ion observables are based on a fluid-dynamical description, which takes initial densities and flow velocities of the produced QCD matter as initial conditions. One either parametrizes these initial conditions \cite{Novak:2013bqa, Bernhard:2016tnd, Bass:2017zyn, Moreland:2018gsh, Bernhard:2019bmu, JETSCAPE:2020mzn, Nijs:2020roc,Parkkila:2021tqq, Parkkila:2021yha, Nijs:2023yab,Auvinen:2017fjw, JETSCAPE:2023nuf, Soeder:2023vdn} or tries to compute them from a QCD dynamical model for the initial production of gluons and quarks \cite{Niemi:2015qia, Gale:2012rq,Auvinen:2017fjw,Hirvonen:2022xfv}. In both cases there is some number of fit parameters that characterize the initial states, and these will obviously be correlated with the actual QCD-matter properties extracted from the data via  Bayesian inference. It is therefore important to model the QCD-matter initial states based on QCD dynamics as far as is possible, in order to understand the dominant particle production mechanism, to reduce the uncertainties in the extraction of the initial states, and to have predictive power for moving from one system to another.

The EKRT (Eskola-Kajantie-Ruuskanen-Tuominen) model \cite{Eskola:1999fc,Eskola:2000xq,Paatelainen:2013eea,Niemi:2015qia}, which treats the nuclear collisions as collisions of parton clouds, and supplements a perturbative QCD (pQCD) calculation for the production of few-GeV partons (minijets) \cite{Kajantie:1987pd,Eskola:1988yh} with a collinear factorization -inspired QCD saturation mechanism \cite{Paatelainen:2013eea,Niemi:2015qia} for regulating the small-$p_T$ minijet production ($p_T$ being transverse momentum), is an example of such a QCD-based initial state modeling with predictive power. The event-by-event (EbyE) version of the model, EbyE-EKRT \cite{Niemi:2015qia}, has been quite successful in explaining a large collection of heavy-ion bulk observables at the LHC and RHIC \cite{Niemi:2015qia, Niemi:2015voa, Eskola:2017bup, Niemi:2018ijm, Auvinen:2020mpc, Hirvonen:2022xfv}. The latest progress here is the novel MC-EKRT event generator (MC for Monte Carlo), introduced recently in Ref.~\cite{Kuha:2024kmq}, and employed in the present article.

The new features in MC-EKRT \cite{Kuha:2024kmq} relative to EbyE-EKRT \cite{Niemi:2015qia} are that now the produced partonic system contains local fluctuations of the minijet multiplicity, which in turn induce dynamical fluctuations to the saturation controlling the initial parton production. Also per-nucleon conservation of energy and valence-quark numbers are accounted for. MC-EKRT also introduces a new type of spatially dependent nuclear parton distribution functions (snPDFs) that are specific to the nucleon configuration in each event and can cope with the largest density fluctuations of the nucleon densities. Thanks to these new features, MC-EKRT gives initial conditions for full 3+1 D EbyE fluid-dynamics, and thus enables the studies of rapidity-dependent observables, such as rapidity distributions of yields and flow coefficients of charged hadrons in Pb+Pb collisions at the LHC and at the highest-energy Au+Au collisions at RHIC -- see Ref.~\cite{Kuha:2024kmq}.

In this paper, we employ the new MC-EKRT framework for computing event-by-event initial conditions for 2+1 D dissipative shear- and bulk-viscous second-order transient fluid-dynamics in the mid-rapidity unit of 5.02 and 2.76 TeV Pb+Pb collisions at the LHC. In particular, we study the sensitivity of the flow coefficients $v_n$ to the model details, such as the nucleonic width and substructure, the Gaussian smearing in coupling the individual minijets to continuous fluid dynamics, as well as the order in which we do the minijet filtering based on saturation and conservation of energy. In addition, we show how the added minijet multiplicity fluctuations are the piece formerly missing from EbyE-EKRT in explaining the behaviour of the charged multiplicity distributions in the most central collisions. The recently developed neural networks for predicting flow observables directly from the initial energy density event-by-event \cite{Hirvonen:2023lqy, Hirvonen:2024ycx}, are also utilized. As the main result of this paper, we show that a detailed simultaneous description of the $v_n$'s requires saturation to be the driving QCD mechanism for initial parton production. In particular, this result calls for further nucleonic substructure -- hotspots -- to be introduced in MC-EKRT. We also implement these in MC-EKRT and discuss their interesting interplay with saturation, in describing the $v_2/v_3$ ratio as well as in explaining the measured charged multiplicity distributions.
\vspace{-0.6cm}

\section{MC-EKRT INITIAL STATE FOR FLUID DYNAMICS}
\vspace{-0.3cm}
\subsection{Minijet sampling}
\vspace{-0.3cm}

The MC-EKRT event generator of Ref.~\cite{Kuha:2024kmq} produces partonic initial states, i.e.\ saturated systems of gluons and quarks with $p_T\gtrsim p_0\sim 1$~GeV, that can be fed as initial conditions to 3+1 D event-by-event fluid-dynamical simulations. The generation of such MC-EKRT initial states proceeds via the following steps (for details, see Ref.~\cite{Kuha:2024kmq}):

First, the nucleon configurations of the colliding (here spherically symmetric) nuclei $A$ and $B$ are generated by sampling the standard 2-parameter Woods-Saxon distribution, and by requiring an exclusion radius of 0.4~fm. A squared impact parameter $b_{AB}^2$ for the $A$+$B$ collision, defining the distance between the centers of masses of the colliding nuclei, is sampled from a uniform distribution. In the absence of hotspots (i.e.\ without sub-nucleonic density fluctuations), the $A$+$B$ collision is triggered using MC Glauber-like black-disc nucleons with a trigger cross section identical to the inelastic nucleon-nucleon cross section $\sigma_\text{inel}^{NN}$, which is obtained from the measured total and elastic nucleon-nucleon cross sections as a function of the nucleon-nucleon center-of-momentum (cms) system energy $\sqrt{s_{NN}}$ \cite{COMPETE:2002jcr, TOTEM:2017asr}.

Once the  $A$+$B$ collision is triggered, MC-EKRT does not consider nucleonic sub-collisions at all but pictures the entire nuclear collision as a collision of two extensive parton clouds. For distributing the parton sub-clouds spatially around each nucleon, MC-EKRT assumes a Gaussian thickness function,
\begin{equation}
T_{N}(\bar{s})=\frac{1}{2\pi\sigma_N^2} \exp\left(-\frac{|\bar s|^2}{2\sigma_N^2}\right), \label{E:TN}
\end{equation}
with a width parameter $\sigma_N=\sigma_N(\sqrt{s_{NN}})$ that is obtained from exclusive photo-production of $J/\Psi$ in photon-proton collisions at HERA~\cite{ZEUS:2002wfj,Eskola:2022vpi}.
Then, multiple dijet production, i.e. the number of independent dijets with jet transverse momentum $p_T\ge p_0=1$~GeV, that is assigned to originate from each $ab$ pair, is sampled from a Poissonian probability distribution with a mean
\begin{equation}
\bar N_\text{jets}^{ab}
= T_{NN}(\bar{b}_{ab})\, \sigma^{ab}_{\text{jet}}(p_0,\sqrt{s_{NN}},\{\bar{s}_a\},\{\bar{s}_b\}), \label{E:barN}
\end{equation}
where $T_{NN}(\bar{b}_{ab})$ is the nucleonic overlap function and $\bar{b}_{ab}$ is the impact parameter between the nucleons $a$ and $b$, while $\sigma^{ab}_{\text{jet}}$ is the integrated pQCD (mini)jet cross section, which MC-EKRT computes using the novel snPDFs for $a$ and $b$, and all possible leading-order (LO) partonic $2\rightarrow 2$ sub-processes. A cms-energy dependent multiplicative $K$-factor is introduced to $\sigma^{ab}_{\text{jet}}$  as a free fit parameter, to account for the missing higher order contributions. The (mini)jet cross section depends on the transverse momentum cut-off parameter $p_0$, on the cms energy $\sqrt{s_{NN}}$, as well as on the transverse locations $\bar s_a$ and $\bar s_b$ of $a$ and $b$ in the nucleon-configurations of $A$ and $B$, indicated here with  $\{\bar s_a\}$ and $\{\bar s_b\}$.

As explained in detail Ref.~\cite{Kuha:2024kmq}, the novel snPDFs are now nucleon-configuration specific and  account for the nuclear modifications of each nucleon's PDFs caused by all other nucleons in the nucleus. In other words, the MC-EKRT snPDFs are nucleon-specific and nucleon-configuration specific.  Also noteworthy is that these novel snPDFs can fully cope with the event-by-event density fluctuations, which was not the case with the formerly developed spatial nPDFs, such as those in Ref.~\cite{Helenius:2012wd}. The MC-EKRT snPDFs are normalized (averaging over all nucleons in each nucleus and over a large number of nuclei) to the spatially averaged nuclear PDF modifications of the EPS09LO set \cite{Eskola:2009uj}, and CT14LO \cite{Dulat:2015mca} are employed for the free proton PDFs.

Finally, the transverse location for each produced dijet is sampled from the product of the two overlap functions $T_N$,  whose transverse integral gives the usual overlap function $T_{NN}$. The kinematic variables and the flavor chemistry of the produced partons, along with identifying the valence quark-consuming processes, is sampled from the differential jet sub-cross sections, as explained in Ref.~\cite{Kuha:2024kmq}.

\subsection{Minijet filtering}

The next, and decisive, step in MC-EKRT is the filtering of the excessive candidate-dijets, based on the EKRT saturation \cite{Eskola:1999fc,Eskola:2000xq,Paatelainen:2013eea,Niemi:2015qia} and conservation of energy and valence quark numbers. As explained in \cite{Paatelainen:2013eea,Niemi:2015qia,Kuha:2024kmq} saturation here is expected to occur when all the higher-order $(n>2)\rightarrow 2$ parton processes start to dominate over the $2\rightarrow 2$ ones. For maintaining collinear factorization at the highest values of jet transverse momenta, the filterings are performed in the order of decreasing factorization scale, which here is the jet $p_T$. Then, the highest-$p_T$ partons can remain in the system while the lower-$p_T$ ones may get filtered away.

For the saturation filtering, MC-EKRT assigns a transverse radius $1/(\kappa_\text{sat} p_T)$ for each dijet candidate, where $\kappa_\text{sat}$ is a packing factor, a free parameter to be fitted from the data. The transverse position of each candidate dijet is kept track of, and a candidate dijet gets filtered away if it overlaps with any of the previously accepted dijets. As shown in Ref.~\cite{Kuha:2024kmq}, after the saturation filtering the $p_T$ distribution of surviving partons is not anymore sensitive to the original cut-off parameter $p_0$ but now saturation is the dynamical and local regulation mechanism for these distributions. This is the major difference to the traditional minijet eikonal models (and models alike)  which are employed in event generators describing multiparton interactions, such
as HIJING \cite{Wang:1991hta}.

Similarly, MC-EKRT keeps track of all the longitudinal momentum fractions and valence quarks drawn out from their mother nucleons by the candidate dijets. If the candidate dijet would make its mother nucleon exceed its energy or valence-quark budget, again checking the dijet candidates in the order of decreasing $p_T$, then that dijet candidate gets filtered away. In the EKRT framework, in the spirit of suggesting saturation as the dominant QCD-mechanism that regulates and controls initial parton production in highest-energy nuclear collisions, the default is to do the saturation filtering first, and only then the energy and valence-quark number conservation filterings. There is, however, an option in the code which we utilize and consequences we study in this paper, to have all the filterings done simultaneously.

\subsection{Nucleon substructure and hotspot trigger}
\label{sec:substructure}
The fluctuating substructure to the nucleons of the MC-EKRT framework is implemented as follows. While there is clear evidence that the nucleon substructure is necessary for describing the measured incoherent $J/\psi$ photo-production~\cite{Mantysaari:2016ykx}, the situation is less clear in heavy-ion collisions. The global analyses performed in Refs.~\cite{Moreland:2018gsh, Nijs:2023yab} provide a slight preference towards the inclusion of the nucleon substructure, but the evidence is not conclusive. However, these analyses use the T$_\text{R}$ENTo \cite{Moreland:2014oya} initial state model, in which the effect of the substructure can partly be compensated with other initial state parameters.

In the MC-EKRT model, the addition of the nucleon substructure enhances the saturation effects since it confines the minijet production into more localized transverse regions. This leads to a change in the initial geometry, which might have an impact on the flow observables. The nucleon substructure is implemented by introducing Gaussian hotspots to the nucleon thickness function:
     \begin{equation}
     \label{eq:hs_Ta}
        T_N(\bar{s}) = \frac{1}{N_h}\sum_{i=1}^{N_h}  \frac{1}{2\pi\sigma_h^2}\exp\left(-\frac{\abs{\bar{s}-\bar{s}^h_i}^2}{2\sigma_h^2}\right),
     \end{equation}
where $N_h$ is the number of hotspots, and $\sigma_h$ is the width of the hotspot. In this article, $N_h = 3$ is always used when the nucleon substructure is enabled. The hotspot locations $\bar{s}^h_i$ are sampled from a 2-dimensional Gaussian distribution with a width $\sigma_s$. The total nucleon width $\sigma_N$ is then related to the hotspot widths via $\sigma_N^2 = \sigma_s^2 + \sigma_h^2$. Therefore, only two of the three widths are independent.
As in Refs.~\cite{H1:2005dtp, Eskola:2022vpi, Kuha:2024kmq}, the energy dependence of the total nucleon width is parametrized as $\sigma_N = \sqrt{b}$ with
\begin{equation}
b/\text{GeV}^{-2} = b_0 + 4\alpha'_P\log\left(\frac{{W}}{W_0}\right). \label{E:b_param}
\end{equation}
where $W=\sqrt{s_{NN}}$, and $b_0$, $\alpha'_P$ and $W_0$ are fit parameters. In the present paper, our default choice of parameters, based on the H1 measurements~\cite{H1:2005dtp}, are $b_0=4.63$, $\alpha'_P=0.164$ and $W_0=90\;\text{GeV}$. This corresponds to $\sigma_N = 0.517$~fm for 2.76 TeV, and $\sigma_N = 0.532$~fm for 5.023 TeV collision energies.

In principle, the nucleon substructure needs to be accounted for when performing the triggering of the nuclear collision event~\cite{Loizides:2016djv, Bozek:2019wyr} since otherwise there might be events where the collision is accepted even though there is no hadronic interaction. As mentioned before, without any substructure, the triggering is done by assuming hard-sphere scattering between two nucleons.
The event is accepted if the distance $d_{\rm min}^{NN}$ between any nucleons $a \in A$ and $b \in B$ satisfy
        \begin{equation}
        \label{eq:trigger}
            d^{NN}_{\rm min} < \sqrt{\frac{\sigma_{\text{inel}}^{{ NN}}}{\pi}},
        \end{equation}
where $\sigma_{\text{inel}}^{{ NN}}$ is the inelastic nucleon-nucleon cross section. The same kind of geometrical criterion can be extended to account for the locations of the hotspots. That is, the triggering with the nucleon substructure is done based on the minimum distance between two colliding hotspots $ d^{HS}_{\rm min}$, i.e.
        \begin{equation}
        \label{eq:trigger_hs}
            d^{HS}_{\rm min} < \sqrt{\frac{\sigma_{{HS}}}{\pi}},
        \end{equation}
where $\sigma_{{HS}}$ is an effective hotspot-hotspot cross section fitted to reproduce the same nucleus-nucleus cross section as obtained with condition~\eqref{eq:trigger}. Therefore, the value of $\sigma_{{HS}}$ will depend on the hotspot sampling-width $\sigma_s$ and the collision system.

Even though in principle hotspot triggering could have a notable impact, we have noticed that in most cases all the measured observables remain nearly unchanged in the 0-80\% centrality range. The largest effects are most visible in the most peripheral charged particle multiplicity region, where usually no measured data are given. In the 60-80\% centralities, the differences in charged particle multiplicities are only a few percent at most. However, since in MC-EKRT we sample dijets from the same nucleon configuration until at least one is produced in a collision, the addition of hotspot triggering there speeds up the generation of the initial states.

\subsection{Initialization of fluid dynamics}
The initial condition of fluid dynamics is the energy-momentum tensor $T^{\mu\nu}$ at some initial proper time $\tau_0$. However, MC-EKRT produces a list of massless partons with known momentum rapidities $y_i$, transverse momenta $\bold{p}_{Ti}$, and transverse coordinates $\bold{x}_{\perp, 0i}$. Thus, the partons need to be propagated to the $\tau_0$ surface and converted to the components of the energy-momentum tensor.  Here we assume that all the partons are produced at the longitudinal location $z_i = 0$ at time $t = 0$, and that they propagate as free particles to the proper time $\tau_0 = 0.2$~fm. Therefore, spacetime and momentum rapidities are equivalent, i.e.\ $\eta_{s, i} = y_i$. The spacetime coordinates of the parton $i$ are then $(\tau_0, \bold{x}_{\perp i}(\tau_0), \eta_{s, i})$ where $\bold{x}_{\perp i}(\tau_0) = \bold{x}_{\perp, 0i} + \tau_0 \bold{p}_{Ti}/p_{Ti}$.

The components of the energy-momentum tensor in the $\tau-\eta_s$ coordinates are obtained as in Ref.~\cite{Kuha:2024kmq},
\begin{align}
\label{eq:energymomentum_ebye}
 &T^{\alpha\beta} (x^\alpha)  = \sum_{i}\int d^2\mathbf{p}_T dy \frac{p^{\alpha}p^{\beta}}{p^\tau}\frac{1}{\tau} \cosh(y - \eta_{s})  \\
 & \times \delta^{(2)}(\mathbf{x}_\perp - \mathbf{x}_{\perp i})\delta(\eta_s - \eta_{s, i}) \delta^{(2)}(\mathbf{p}_T - \mathbf{p}_{T i}) \delta(y - \eta_{s}), \notag
\end{align}
where the four-momentum $p^\alpha= (p^\tau, \bold{p}_T, p^\eta)$ at a spacetime location $x^\alpha = (\tau, \bold{x}_\perp, \eta_s)$ is given by
    \begin{equation}
    \label{eq:FourMomentum_hyperbolic}
            p^\alpha  = \left(
            \begin{matrix}
                p_{T} \cosh(y - \eta_{s}) \\
                \mathbf{p}_T \\
                \tau^{-1} p_{T} \sinh(y -\eta_{s})
            \end{matrix}
            \right).
    \end{equation}

Depositing all energy and momentum of a parton into a single cell on a hydro grid as suggested by the delta functions appearing in Eq.~\eqref{eq:energymomentum_ebye} would lead to extreme fluctuations in energy and momentum densities. To obtain smooth density distributions, smearing is required. Here we are performing 2+1 D hydrodynamic simulations, where a natural choice is to let all partons that are produced in the midrapidity window $\Delta y$ contribute to the fluid dynamical initial state. That is, in Eq.~\eqref{eq:energymomentum_ebye} we replace $\delta(\eta_s - \eta_{s,i}) \to \theta(\Delta y/2 - \abs{\eta_{s,i}})/\Delta y$, where $\theta$ is the Heaviside theta function. Here we use $\Delta y = 1.0$, but we have tested that the final results are practically insensitive to the choice of $\Delta y$ as long as $0.5 \leq \Delta y \leq 2.0$. The smearing in the transverse $(x,y)$ plane is performed by replacement $\delta^{(2)}(\bold{x}_\perp - \bold{x}_{\perp i}) \to g_\perp (\bold{x}_\perp; \bold{x}_{\perp i})$, where
\begin{align}
\label{eq:SmearingInitial_trans}
    g_\perp (\mathbf{x}_\perp; \mathbf{x}_{\perp i}) = \frac{C_\perp}{2\pi \sigma_\perp^2}  \exp \left[ -\frac{(\mathbf{x}_\perp - \mathbf{x}_{\perp i})^2}{2\sigma_\perp^2 } \right]
\end{align}
is a Gaussian distribution with transverse smearing width $\sigma_\perp$ which is normalized as
 \begin{align}
    \label{eq:SmearingNormInitial_2D}
    \int d^2\mathbf{x}_\perp g_\perp (\mathbf{x}_\perp; \mathbf{x}_{\perp i})  = 1.
\end{align}
The computation cost is reduced by imposing a $\pm 3\sigma_\perp$ cut-off on the smearing range, and the coefficient $C_\perp$ takes care of the unit normalization.

As in Ref.~\cite{Kuha:2024kmq}, we only consider the local rest frame energy density $e$ when initializing the fluid dynamical system, i.e.\ we neglect the initial transverse velocity and the initial components of the shear-stress tensor. Therefore, the initialization is determined by
\begin{align}
\label{E:EMtensor_initialization_pt_2D}
   T^{\tau \tau} & (\tau_0, \mathbf{x}_\perp, \Delta y) = \notag \\
   & \frac{1}{\tau_0 \Delta y}\sum_i p_{T i} g_\perp (\mathbf{x}_\perp; \mathbf{x}_{\perp i}) \theta(\Delta y/2-\abs{y_i}),
\end{align}
which in this case coincides with $e$. The remaining components are then obtained, using the equation of state, as $T^{ij} = P(e)\delta^{ij}$.

Finally, we emphasize that even if we utilize only the midrapidity minijets in computing the above initial conditions, the underlying MC-EKRT event generation is fully 3~D. Thus, the midrapidity initial conditions are influenced also by the finite-rapidity effects in saturation and in energy conservation.

\section{Fluid simulation framework}
The simulations performed in this article focus on midrapidity observables and therefore we assume that the longitudinal expansion of the system is boost invariant. The same framework as in Ref.~\cite{Hirvonen:2022xfv} is used, i.e.\ we evolve the initially formed strongly interacting matter using dissipative fluid dynamics, and compute the final particle spectra at the dynamical decoupling surface. Additionally, the neural networks trained in Ref.~\cite{Hirvonen:2023lqy} are utilized for significantly decreasing the computation time of the simulations. In this section, we give a brief recapitulation of each aspect of the framework.
\subsection{Fluid dynamics}
Fluid dynamics is based on the local conservation laws for energy, momentum, and conserved charges. Here we neglect the conserved charges, in which case the conservation law for the energy-momentum tensor, $\partial_\mu T^{\mu\nu} = 0$, controls the dynamics. The energy-momentum tensor can be decomposed with respect to 4-velocity $u^\mu$ as
\begin{equation}
T^{\mu\nu} = e u^\mu u^\nu - P \Delta^{\mu\nu} + \pi^{\mu\nu},
\label{eq:energymomentum}
\end{equation}
where $\Delta^{\mu\nu} = g^{\mu\nu} - u^\mu u^\nu$ is a projection operator, $P = -\frac{1}{3}\Delta_{\mu\nu}T^{\mu\nu} $ is the total isotropic pressure, $e = T^{\mu\nu} u_\mu u_\nu$ is the local rest frame energy density, and $\pi^{\mu\nu} = T^{\langle \mu\nu \rangle}$ is the shear-stress tensor. The angular brackets denote the symmetric, traceless part of the tensor that is orthogonal to the fluid 4-velocity. Here the fluid velocity is defined in the Landau frame, i.e. $T^\mu_{\,\,\nu} u^\nu = e u^\mu$. The bulk viscous pressure is defined as the deviation of the isotropic pressure $P$ from the equilibrium pressure $P_0$, i.e. $\Pi = P -P_0$. The equilibrium pressure is given by the equation of state (EoS) of the QCD matter at zero baryon density, $P_0 = P_0(e)$. In this work, we use the $s95p$-v1 parametrization~\cite{Huovinen:2009yb} for the EoS, which includes the partial chemical decoupling at $T_{\rm chem} = 155$ MeV. The partial chemical decoupling is implemented by adding temperature-dependent chemical potentials for each hadron in the hadronic part of the EoS~\cite{Bebie:1991ij, Hirano:2002ds, Huovinen:2007xh}.

The conservation laws together with the EoS are enough to solve the evolution in equilibrium, but additional constraints are needed when dissipative effects are present. The dissipative parts of the energy-momentum tensor are the shear-stress tensor and the bulk viscous pressure. In the formalism by Israel and Stewart~\cite{Israel:1979wp}, the equations of motion for dissipative parts take a form
\begin{equation}
 \tau_\Pi \frac{d}{d\tau}\Pi+ \Pi = -\zeta \theta  - \delta_{\Pi\Pi}\Pi\theta + \lambda_{\Pi\pi}\pi^{\mu\nu}\sigma_{\mu\nu},
\label{eq:IShydrobulk}
\end{equation}

\begin{eqnarray}
 \tau_\pi \frac{d}{d\tau}\pi^{\langle \mu \nu \rangle} + \pi^{\mu\nu}  =& 2\eta \sigma^{\mu\nu}  + 2 \tau_\pi
 \pi_\alpha^{\,\left\langle \mu\right.}\omega_{\vphantom{12pt}}^{\left.\nu\right\rangle \alpha} \notag \\
 & - \delta_{\pi\pi} \pi^{\mu\nu} \theta-\tau_{\pi\pi} \pi_{\alpha}^{\langle \mu} \sigma_{\vphantom{12pt}}^{\nu\rangle \alpha} \\
 &+ \varphi_7 \pi_\alpha^{\langle \mu} \pi_{\vphantom{12pt}}^{\nu\rangle \alpha}+\lambda_{\pi\Pi} \Pi \sigma^{\mu\nu} \notag,
\end{eqnarray}
where $\theta = \nabla_{\mu}u^{\mu}$ is the expansion rate, $\sigma^{\mu\nu} = \nabla^{\langle \mu}u^{\nu\rangle}$ is the strain-rate tensor, and $\omega^{\mu\nu} = \frac{1}{2}\left(\nabla^{\mu}u^{\nu} - \nabla^{\nu}u^{\mu}\right)$ is the vorticity tensor. The first-order transport coefficients $\eta$ and $\zeta$ are called shear and bulk viscosity respectively. In a 14-moment approximation to the massless gas~\cite{Denicol:2010xn, Denicol:2012cn, Molnar:2013lta, Denicol:2014vaa}, the first-order transport coefficients are related to the shear and bulk relaxation times as
\begin{equation}
 \tau_\pi = \frac{5\eta}{e+P_0}, \hspace{3mm} \tau_\Pi = \left(\mathrm{15} \Big(\frac{1}{3}-c_s^2\Big)^2(e+P_0)\right)^{-1}\zeta,
 \label{eq:relaxation_time}
\end{equation}
and the remaining second-order transport coefficients are
\begin{equation}
    \begin{split}
        &\delta_{\Pi\Pi} = \frac{2}{3}\tau_\Pi, \hspace{2mm} \lambda_{\Pi\pi} = \frac{8}{5}\Big(\frac{1}{3}-c_s^2\Big)\tau_\Pi, \hspace{2mm} \delta_{\pi\pi} = \frac{4}{3}\tau_\pi, \\
        &  \tau_{\pi\pi} = \frac{10}{7} \tau_\pi, \hspace{2mm} \varphi_7 = \frac{9}{70 P_0}, \hspace{2mm} \lambda_{\pi\Pi} = \frac{6}{5} \tau_\pi,
    \end{split}
\end{equation}
where $c_s$ is the speed of sound. The specific shear viscosity $\eta/s$ and specific bulk viscosity $\zeta/s$ are from the $\eta/s = dyn$ parametrization introduced in Ref.~\cite{Hirvonen:2022xfv}.

\subsection{Decoupling and particle spectra}
The fluid dynamic evolution is continued until reaching the kinetic decoupling surface. Here the decoupling surface is determined by the dynamical decoupling conditions
\begin{eqnarray}
 \mathrm{Kn} &=& \tau_{\pi} \theta = C_{\mathrm{Kn}} \\
 \label{eq:freeze1}
 \frac{\gamma\tau_{\pi}}{R} &=& C_{R},
 \label{eq:freeze2}
\end{eqnarray}
where $\mathrm{Kn}$ is the Knudsen number, $\gamma$ is the Lorentz gamma factor, and the coefficients $C_{\mathrm{Kn}}$ and $C_{R}$ are proportionality constants of $\mathcal{O}(1)$ which are fitted to the measured data. Here, values $C_{\mathrm{Kn}} = 0.8$ and $C_{R}= 0.15$ are used according to Ref.~\cite{Hirvonen:2022xfv}. The size of the system $R$ is defined as
\begin{equation}
    R=\sqrt{\frac{A}{\pi}},
\end{equation}
where $A$ is the area in the transverse $(x, y)$ plane where $\mathrm{Kn} < C_{\mathrm{Kn}}$. Additionally, the decoupling is forced to happen in the hadronic phase of the QCD matter, i.e. when $T<150$ MeV. Given these conditions the decoupling surface is determined using the Cornelius algorithm~\cite{Huovinen:2012is}.

At the decoupling surface $\Sigma$ with the directed surface element $d\Sigma_\mu$, the Lorentz-invariant particle spectrum for particle type $i$ is computed according to the Cooper-Frye integral,
    \begin{equation}
        E \frac{d^3N_i}{d^3k} = \int_\Sigma d\Sigma_\mu k^\mu f_i(x, k),
        \label{eq:Cooper-Frye}
    \end{equation}
where $E$ and $k^\mu$ are particles energy and 4-momentum, respectively. The distribution function for particle species $i$ is decomposed into in- and out-of-equilibrium parts as $f_i = f_{0i}+\delta f_i$, where the equilibrium part is given by

    \begin{equation}
        f_{0i}(x, k) = \Big[\exp\Big(\frac{k_i^\mu u_\mu - \mu_i}{T}\Big)\pm 1 \Big]^{-1},
    \end{equation}
where $+$($-$) sign is for fermions (bosons), and $\mu_i$ is the chemical potential. Here, the viscous corrections to the equilibrium distribution are of the form~\cite{Hosoya:1983xm, Gavin:1985ph, Sasaki:2008fg, Bozek:2009dw}
\begin{equation}
\begin{split}
\delta f_i =& -f_{0i}\tilde{f}_{0i} \frac{C_{bulk}}{T}\bigg[\frac{m^2}{3 E_k}-\Big(\frac{1}{3}-c_s^2\Big) E_k \bigg] \Pi \\
&+\frac{f_{0i}\tilde{f}_{0i}}{2T^2(e + P_0)} \pi^{\mu\nu} k_{\mu} k_{\nu},
\end{split}
\end{equation}
with $\tilde{f}_{0i}=1 \pm f_{0i}$ ($+$ for bosons and $-$ for fermions) and the coefficient
\begin{equation}
\frac{1}{C_{bulk}}= \sum_i \frac{g_i m_i^2}{3T} \int \frac{\mathrm{d}^3\mathbf{k}}{(2\pi)^3 k^0} f_{0i}\tilde{f}_{0i}\bigg[\frac{m_i^2}{3E_k}-\Big(\frac{1}{3}-c_s^2\Big)E_k\bigg],
\end{equation}
where $g_i$ is the degeneracy factor. After computing the spectra from Eq.~\eqref{eq:Cooper-Frye}, the 2- and 3-body decays of unstable particles are computed as in Ref.~\cite{Sollfrank:1991xm}.

\begin{figure*}
    \centering
    \includegraphics[width = \textwidth]{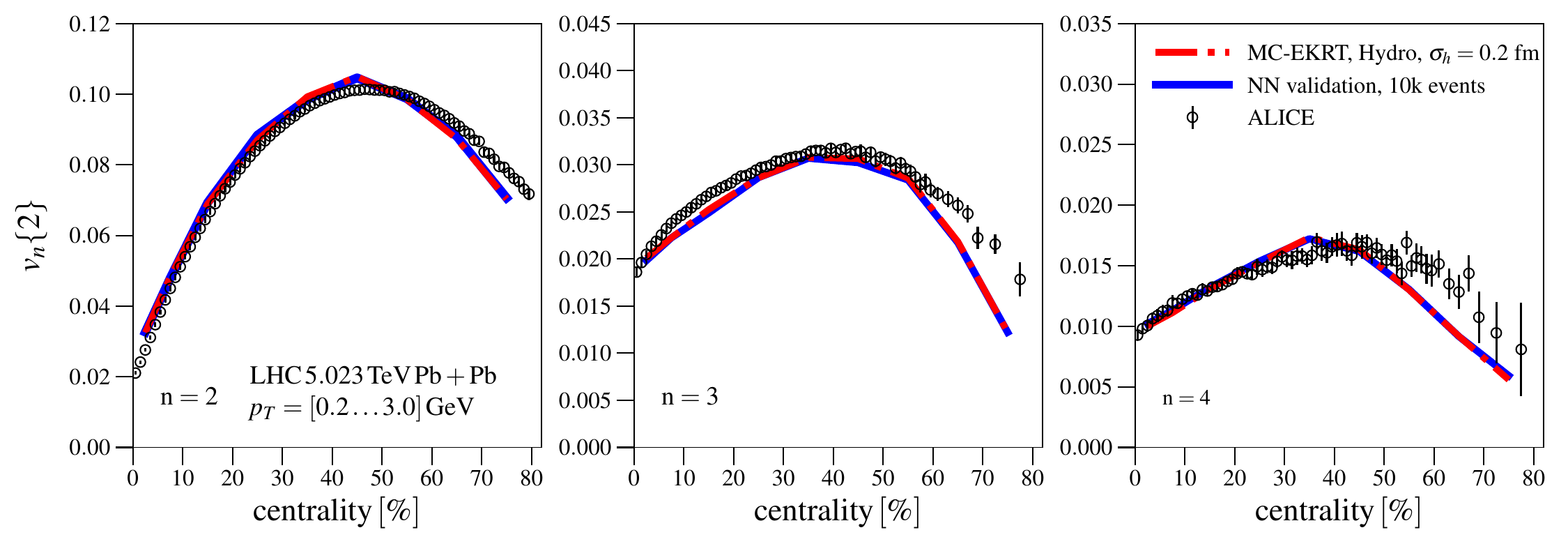}
    \caption{Neural network validation test for the flow coefficients $v_2\{2\}, v_3\{2\}$, and $v_4\{2\}$ in 5.023 TeV Pb+Pb collisions. The networks were trained with the EbyE-EKRT data from Ref.~\cite{Hirvonen:2022xfv} as described in Ref.~\cite{Hirvonen:2023lqy}. The hydro results and the neural network validation results were obtained from 10k MC-EKRT initial states which included hotspots and multiplicity fluctuations that were not present in the training data. The measured data are from the ALICE Collaboration~\cite{ALICE:2018rtz}.  }
    \label{fig:vn_val_MC_EKRT}
\end{figure*}

\subsection{Neural Networks}
\label{sec:nn}
To reduce the computational cost of the simulations, deep convolutional neural networks trained in Ref.~\cite{Hirvonen:2023lqy} are utilized here for predicting final state event-by-event observables at midrapidity. Each neural network takes the discretized initial energy density profile in the transverse-coordinate $(x,y)$ plane as an input, and outputs one $p_T$-integrated observable. Separate neural networks are used to predict flow coefficients $v_n$, charged particle multiplicities $dN_{\rm ch}/d\eta$, and mean transverse momenta $[p_T]$. Predicting flow observables with neural networks is many orders of magnitude faster than performing full hydrodynamic simulations. For example, predicting results for 10 million events takes only around 20 hours with Nvidia Tesla V100 GPU.

As the training data for the neural networks is from Ref.~\cite{Hirvonen:2022xfv}, and the predictions made using these networks emulate the same dynamics as the training data, the viscosities $\eta/s$, and $\zeta/s$, and other parameters affecting the fluid-dynamical evolution are the same ones as in Ref.~\cite{Hirvonen:2022xfv}.

In Ref.~\cite{Hirvonen:2023lqy}, it was demonstrated that the neural networks work accurately when using the EbyE version of the EKRT model. However, it is non-trivial that the accuracy of the neural networks, which are trained by the EbyE-EKRT data from Ref.~\cite{Hirvonen:2022xfv}, and \emph{not} from MC-EKRT, would extend to the MC-EKRT initial states with hotspots, where the initial geometry can be significantly different. Therefore, the neural networks were validated by generating 10k MC-EKRT initial states and comparing the neural network predictions against 2+1 D fluid dynamical simulations for the 5.023 TeV Pb+Pb collision system. The validation tests for the flow coefficients $v_2, v_3$, and $v_4$ are shown in Fig.~\ref{fig:vn_val_MC_EKRT}. The initial state parameters used in the validation test were $\kappa_{\mathrm{sat}} = 2.5$, $K = 2.2$, $\sigma_\perp = 0.4$ fm, and $\sigma_h = 0.2$ fm.  The obtained excellent agreement between the fluid dynamical simulations and neural network predictions illustrates the versatility of the neural networks with different initial conditions. Additionally, we have verified that the accuracy of the employed neural networks remains good for other training observables as well.

\section{Results}
In this section, we present the results of fluid-dynamical simulations with MC-EKRT initial states for midrapidity bulk observables, and compare the results against the earlier EbyE EKRT work~\cite{Hirvonen:2022xfv}. All the fluid dynamical results are generated using our neural networks, and they contain 50k collision events, except the multiplicity distribution results which are obtained from 150k events. As discussed in Sec.~\ref{sec:nn}, the neural network results correspond to the fluid dynamical simulations with the matter properties and decoupling parameters from Ref.~\cite{Hirvonen:2022xfv}. Therefore, any differences between the presented results are due to differences in the initial states.

When examining the effects of the initial state through final state observables, it is important to remember that some observables are highly sensitive to the properties of the matter. For instance, the magnitude of flow coefficients is significantly influenced by the shear viscosity to entropy density ratio $\eta/s$. In contrast, the ratios of flow coefficients are less sensitive to such details, particularly the ratio between $v_3$ and $v_2$, which can provide valuable insights into the geometry and structure of the initial state~\cite{Retinskaya:2013gca}.

\begin{figure*}
    \centering
    \includegraphics[width = \textwidth]{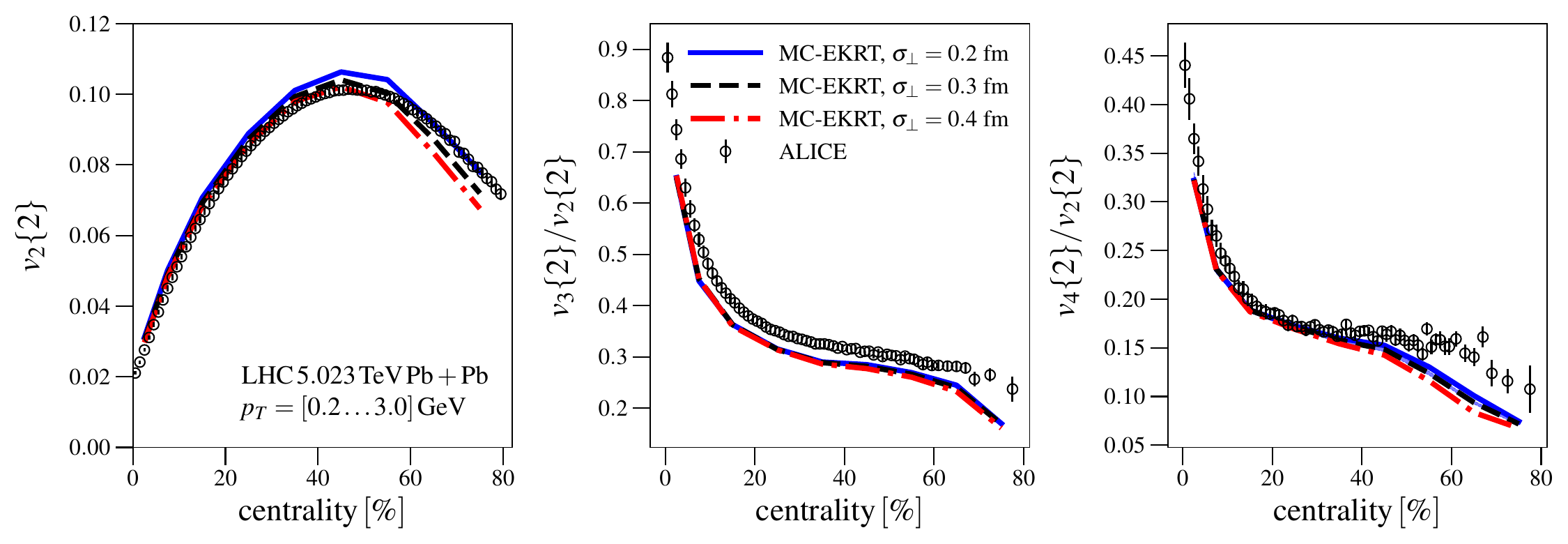}
    \caption{The effect of the Gaussian smearing width $\sigma_\perp$ on the two-particle flow coefficient $v_2\{2\}$ (left panel) and the ratios $v_3\{2\}/v_2\{2\}$ (middle panel), and $v_4\{2\}/v_2\{2\}$ (right panel) in 5.023 TeV Pb+Pb collisions. No nucleon substructure is included here. The experimental data for the ratios are computed based on the ALICE measurements for the two-particle flow coefficients~\cite{ALICE:2018rtz}. }
    \label{fig:vn_smearing_test}
\end{figure*}

\begin{figure*}
    \centering
    \includegraphics[width = \textwidth]{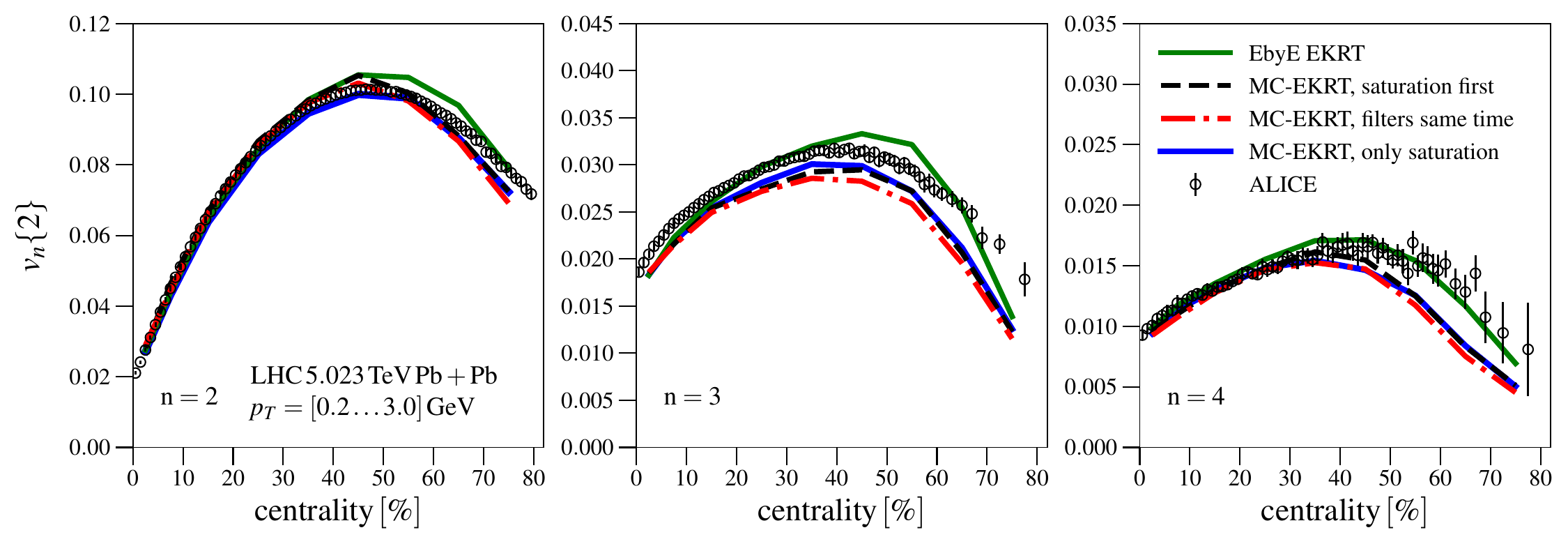}
    \caption{The flow coefficients $v_n\{2\}$ as a function of centrality for 5.023 TeV Pb+Pb collisions. The simulation results with different MC-EKRT filter settings are compared against the ALICE measurements~\cite{ALICE:2018rtz}, and the EbyE-EKRT results from Ref.~\cite{Hirvonen:2022xfv}. No nucleon substructure was included here. }
    \label{fig:vn_filters}
\end{figure*}

The effect of the Gaussian smearing width $\sigma_{\perp}$ is demonstrated in Fig.~\ref{fig:vn_smearing_test}, where the ratios of the flow coefficients $v_2, v_3$, and $v_4$ in 5.023 TeV Pb+Pb collision system are shown as a function of centrality for different smearing widths. The MC-EKRT initial state parameters are set to  $\kappa_{\mathrm{sat}} = 1.4$, and $K = 2.5$. Nucleon substructure is not included in these plots. As can be seen in the left panel, the magnitude of flow is sensitive to the Gaussian smearing width $\sigma_{\perp}$. However, $\sigma_{\perp}$ has only little impact on the ratios between the flow coefficients, as shown by the middle and right panels. Therefore, the parameter $\sigma_{\perp}$ is influencing the flow coefficients in a similar manner as the shear viscosity. Here, and in what follows, we adjust $\sigma_{\perp}$ to obtain the measured $v_2$ in mid-central collisions for all different MC-EKRT results. However, we want to emphasize that this is only done to illustrate the capabilities and uncertainties of MC-EKRT. To get the best overall fit to all different observables, a global analysis is needed, but this is beyond the purpose of this study.
 
An intriguing aspect of the MC-EKRT model is the interplay between the saturation and conservation-law filters. The impact of different filters on the flow coefficients in 5.023 TeV Pb+Pb collisions is illustrated in Fig.~\ref{fig:vn_filters}. In all these scenarios, a value of $K = 2.5$ is used, while the saturation parameter $\kappa_{\mathrm{sat}}$ is adjusted to achieve roughly identical charged particle multiplicities in central collisions. This corresponds to $\kappa_{\mathrm{sat}} = 1.3$ for the saturation-only case, and $\kappa_{\mathrm{sat}} = 1.4$ for the other cases. The nucleon width is set according to the default parametrization from Eq.~\eqref{E:b_param}, i.e.\ $\sigma_N = 0.53$ fm, and no nucleon substructure is introduced. For the saturation-first case $\sigma_\perp = 0.3$ fm, for the case with all filters at the same time
$\sigma_\perp = 0.4$ fm, and for the saturation-only case $\sigma_\perp = 0.3$ fm.

\begin{figure*}
    \centering
    \includegraphics[width = \textwidth]{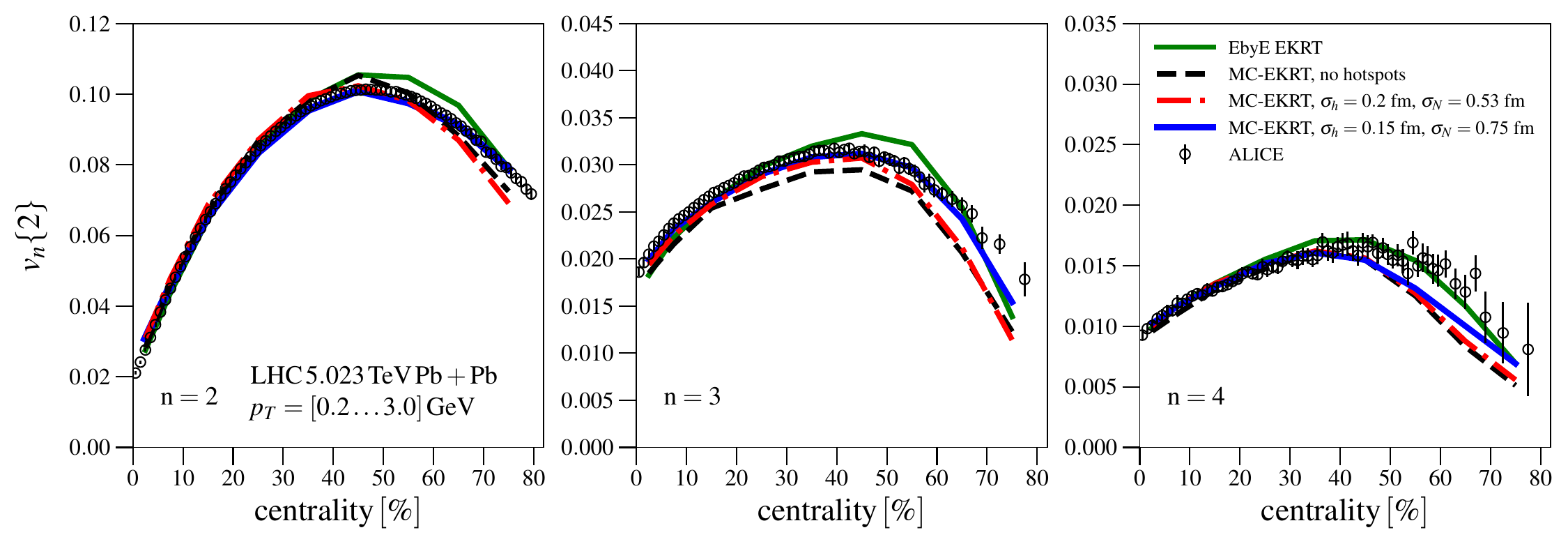}
    \caption{The flow coefficients $v_n\{2\}$ as a function of centrality for 5.023 TeV Pb+Pb collisions. The MC-EKRT results with and without nucleon substructure are compared against the ALICE measurements~\cite{ALICE:2018rtz}, and the EbyE-EKRT results from Ref.~\cite{Hirvonen:2022xfv}.}
    \label{fig:vn_hotspot}
\end{figure*}

\begin{figure*}
    \centering
    \includegraphics[width = \textwidth]{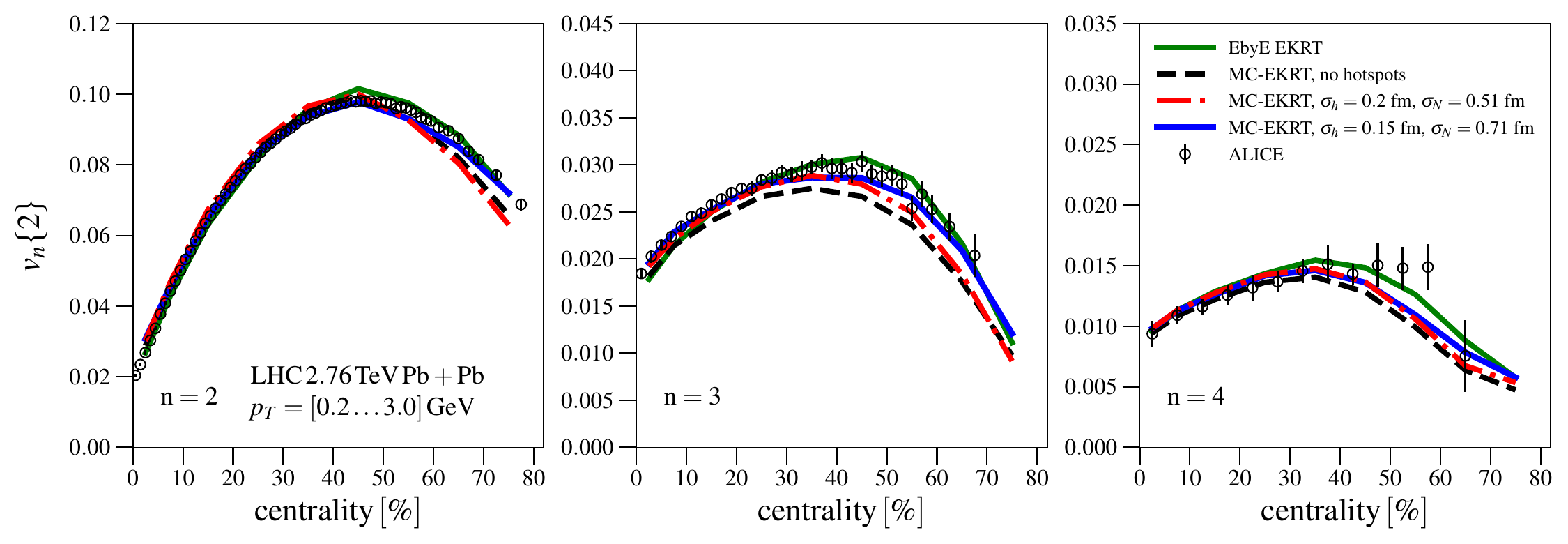}
    \caption{The flow coefficients $v_n\{2\}$ as a function of centrality for 2.76 TeV Pb+Pb collisions. The MC-EKRT results with and without nucleon substructure are compared against the ALICE measurements~\cite{ALICE:2018rtz}, and the EbyE-EKRT results from Ref.~\cite{Hirvonen:2022xfv}.}
    \label{fig:vn_hotspot_LHC2760}
\end{figure*}

The most notable feature in Fig.~\ref{fig:vn_filters} is the significant impact of saturation on the ratio between $v_3$ and $v_2$. The case with only saturation reproduces the measured $v_2$ and $v_3$ most accurately, while the simultaneous application of all the filters leads to a clear underestimation of $v_3$. When saturation is applied before other filters, the results approach those of the saturation-only scenario, as anticipated. The discrepancies in the $v_3/v_2$ ratio arise from the geometrical differences in saturation and momentum conservation. Saturation does not allow geometrical overlap in the transverse plane. This leads to a more evenly distributed energy density profile. Energy conservation, on the other hand, gives no direct geometrical constraints. The stronger the saturation the more the eccentricity $\varepsilon_2$ is suppressed compared to the eccentricity $\varepsilon_3$. The reduced eccentricity $\varepsilon_2$ can be compensated by decreasing the smearing width $\sigma_{\perp}$ so that the elliptic flow $v_2$ remains nearly unchanged, while $\varepsilon_3$ increases. This is reflected in the shown flow coefficients. It is also noteworthy that the $v_3/v_2$ ratio is very similar between the MC-EKRT model with only saturation and the EbyE-EKRT model, which does not explicitly include momentum conservation. Since strong saturation appears to be necessary for matching the measured $v_3/v_2$ ratio, we will now focus exclusively on the scenarios where saturation is applied first, followed by the conservation filters. This approach is also theoretically justified because, in principle, saturation should inherently account for conservation laws. However, achieving this would require implementing saturation through momentum-conserving multiparton distributions to all orders, which is not practically feasible.

Since saturation is sensitive to the nuclear overlap $T_A T_B$ (nuclear thickness function $T_A$ is the sum of $T_N$s), the hotspots introduce interesting dynamics. With the hotspots, $T_A$ can reach $\sim 10$ times higher values than with the average nucleon geometry. Therefore, one would expect the saturation strength and the $v_3/v_2$ ratio to increase when hotspots are included.

The effect of hotspots on the flow coefficients is illustrated in Fig.~\ref{fig:vn_hotspot}, which compares two different hotspot parametrizations. The first parametrization uses the default nucleon width from parametrization Eq.~\eqref{E:b_param}, together with hotspots with width $\sigma_h = 0.2$ fm. In this case, the MC-EKRT parameters are set to $\kappa_{\mathrm{sat}} = 2.5$, $K = 2.2$, and $\sigma_\perp = 0.4$ fm. For the second parametrization, the nucleon width is obtained from Eq.~\eqref{E:b_param}, but this time a significantly stronger energy dependence with $\alpha' = 0.6$ is used. This corresponds to a nucleon width $\sigma_N = 0.75$ fm for 5.023 TeV collision energy. This nucleon width is in line with the many global analyses, where values in the range $\sim 0.6-1.0$ fm are preferred~\cite{Moreland:2018gsh, JETSCAPE:2020mzn, Nijs:2020ors, Nijs:2023yab}. With a wider nucleon, a narrower hotspot with $\sigma_h = 0.15$ fm is used together with parameters $\kappa_{\mathrm{sat}} = 2.5$, $K = 2.4$, and $\sigma_\perp = 0.25$ fm. The saturation-first case from Fig.~\ref{fig:vn_filters} is here left as a reference curve.

As expected, the addition of hotspots appears to increase the $v_3/v_2$ ratio. The best overall fit to the measurements is obtained with the narrow hotspots, i.e. $\sigma_h = 0.15$ fm, corresponding thus to the strongest saturation. In this case, the centrality dependence of $v_2$, and $v_3$ matches nearly perfectly to the ALICE measurements~\cite{ALICE:2018rtz}, while maintaining a good agreement for $v_4$. These findings suggest that the interplay between hotspots and saturation is crucial for the simultaneous description of the flow coefficients and especially of the $v_3/v_2$ ratio.

\begin{figure*}
    \centering
    \includegraphics[width = 0.48\textwidth]{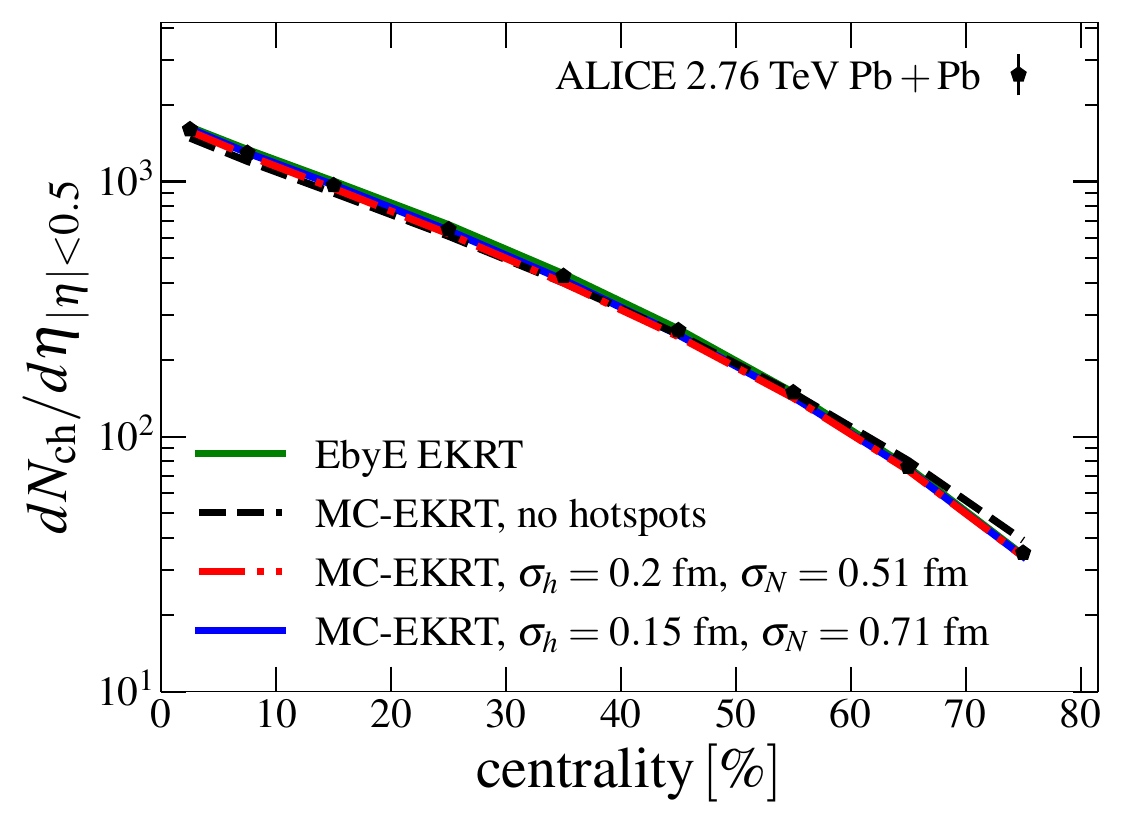}
    \includegraphics[width = 0.48\textwidth]{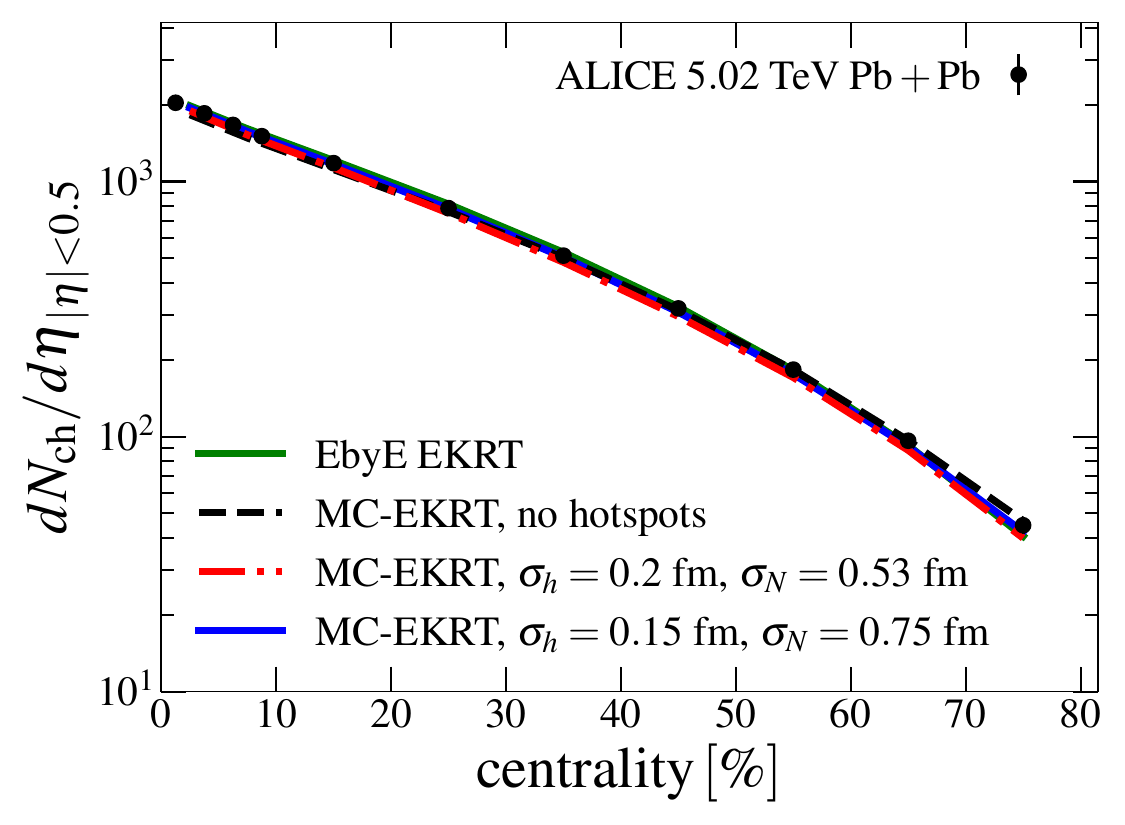}
    \caption{The charged particle multiplicity as a function of centrality in 2.76 TeV (left panel), and 5.023 TeV (right panel) Pb+Pb collisions. The MC-EKRT results with and without nucleon substructure are compared against the ALICE measurements~\cite{ALICE:2010mlf, ALICE:2015juo}, and the EbyE-EKRT results from Ref.~\cite{Hirvonen:2022xfv}. }
    \label{fig:multiplicity_hotspots}
\end{figure*}

In Fig.~\ref{fig:vn_hotspot_LHC2760}, the flow coefficients are shown for 2.76 TeV Pb+Pb collisions. The different curves correspond to the same cases as in Fig.~\ref{fig:vn_hotspot}, but the $K$ factor is adjusted to obtain a reasonable agreement with the measured charged particle multiplicity. The obtained values are $K = 2.5$ for the $\sigma_h = 0.2$ fm case, while the $\sigma_h = 0.15$ fm and the no-hotspots cases both use $K = 2.7$. The agreement between the data and the results is quite similar to the 5.023 TeV collision energy results. At both energies, the narrow-hotspot case with $\sigma_h = 0.15$ fm can describe the measured flow coefficients well, while the centrality dependence of $v_2$ is slightly off for the $\sigma_h = 0.2$ fm case. From Figs.~\ref{fig:vn_hotspot} and \ref{fig:vn_hotspot_LHC2760} it can be seen that MC-EKRT with the nucleon substructure captures the energy dependence of the flow coefficients significantly better than the EbyE-EKRT model.

In Fig.~\ref{fig:multiplicity_hotspots}, the charged particle multiplicity as a function of centrality is shown for the same initial state parametrizations in 2.76 TeV and 5.023 TeV Pb+Pb collisions. The agreement between the results and the ALICE measurements~\cite{ALICE:2010mlf, ALICE:2015juo} is good in all cases, even though there are some minor discrepancies in the centrality behavior. The initial state without hotspots seems to produce slightly too weak a centrality dependence, while, with the hotspots, the centrality dependence is a bit too steep. However, these are small differences, and further improvements could be obtained by fine-tuning the matter properties and initial state parameters.

\begin{figure*}
    \centering
    \includegraphics[width = 0.7\textwidth]{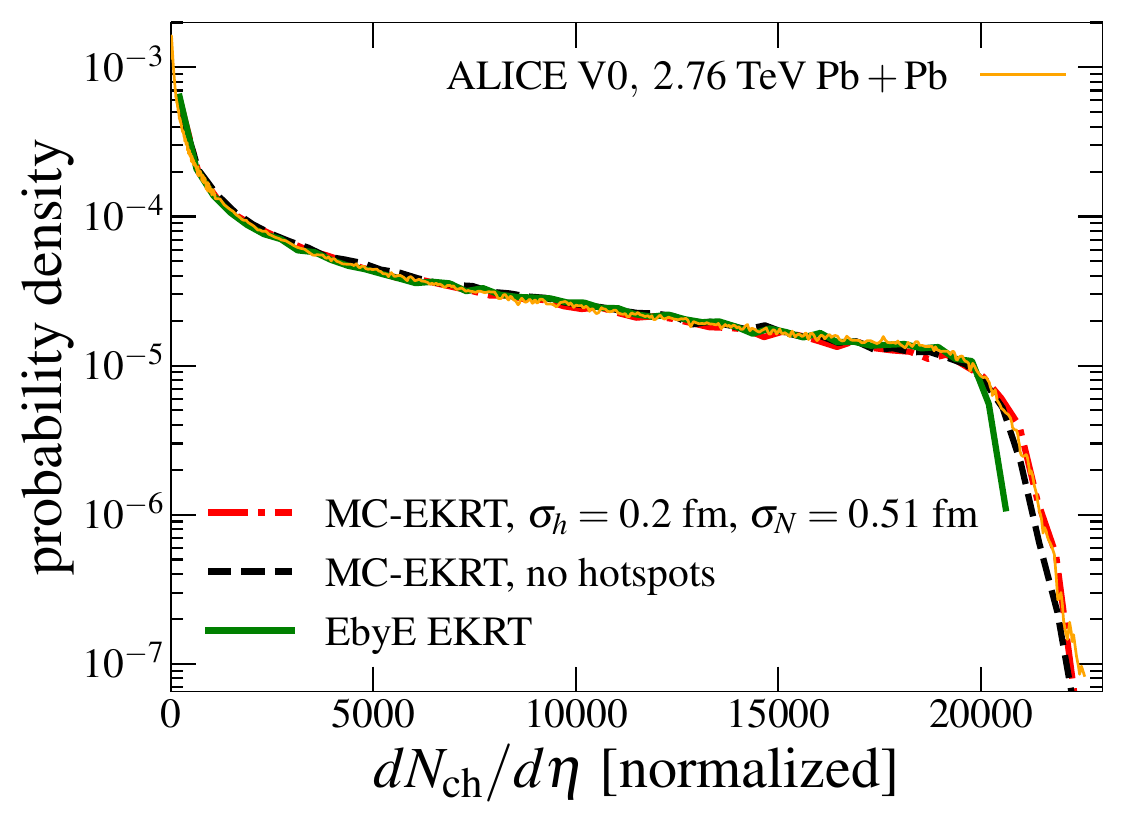}
    \caption{The probability distribution of charged particle multiplicity for 2.76 TeV Pb+Pb collisions. The MC-EKRT results with and without nucleon substructure are compared against the parametrization of the ALICE V0 amplitude read off from Ref.~\cite{ALICE:2013hur}, and the EbyE-EKRT results from Ref.~\cite{Hirvonen:2022xfv}.}
    \label{fig:multiplicity_distribution}
\end{figure*}

The MC-EKRT approach adds minijet-multiplicity originating saturation-scale fluctuations to the EKRT initial state. These fluctuations, together with hotspot fluctuations, should in principle increase the hadron multiplicity fluctuations in the most central collisions. This effect is studied in Fig.~\ref{fig:multiplicity_distribution}, where the charged hadron multiplicity distributions from MC-EKRT with and without hotspots are compared against the EbyE-EKRT results, which do not contain multiplicity-originating fluctuations of the saturation scale or hotspots. To make the results comparable with the V0 amplitude measured by ALICE~\cite{ALICE:2013hur}, they are normalized to have approximately the same mean as the V0 amplitude. As shown also in Ref.~\cite{Niemi:2015qia}, the EbyE-EKRT results almost completely miss the high-multiplicity tail in the distribution. The addition of the further saturation scale fluctuations indeed enhances the high-multiplicity tail in the distribution, and therefore improves the agreement with the measurements as one would expect. The addition of the hotspots is important also for this observable, as it increases the fluctuations and high-multiplicity tail further, leading to a very good agreement with the ALICE data.

\section{Conclusions}
\label{sec:conclusions}

In this article, we have studied the effects of the MC-EKRT initial states on midrapidity flow observables. The computationally slow fluid dynamics simulations were replaced with the neural networks, that could predict flow observables directly from the initial state. The networks used here did not contain any information about the MC-EKRT initial states. Even so, the neural networks did accurately describe the flow observables, emphasizing the versatility and usefulness of the neural networks.

We found that essentially the strength of saturation controls the ratio between two-particle flow coefficients $v_3/v_2$. Without any nucleon substructure, the measured data preferred that no local momentum conservation was enforced, so that the saturation would be the only effect that regulates the initial low-$p_T$ parton production. The addition of the nucleon substructure enhanced the saturation strength, and led to a good agreement with the measured data, even with the local momentum conservation imposed. Our flow coefficient results lend support to having relatively narrow hotspots in a relatively wide nucleon, and rather systematically saturation as the decisive QCD mechanism for regulating the initial parton production.

The results from the MC-EKRT initial state with the nucleon substructure managed to improve the agreement with the LHC measurements relative to the previous EbyE-EKRT model. The novel MC-EKRT model now captures the measured energy dependence of the flow coefficients better, while the added saturation scale fluctuations and the inclusion of hotspots systematically improves the agreement with the measured multiplicity distribution in the most central collisions.

Overall, the MC-EKRT results presented here show an excellent agreement with the data for the flow coefficients and  the charged particle multiplicity. We want to note that this was achieved even without adjusting the QCD matter properties or the dynamical decoupling conditions from previous works, and therefore this acts as a baseline for what can be achieved. More detailed global analysis with more observables and collision systems should be done to constrain the QCD matter properties. Additionally, at the lower collision energies, the finite longitudinal overlap area in the initial collision together with the initial transverse flow can play an important role in the simulations. These aspects were not considered here, but are left as future work.

\acknowledgments

We acknowledge the financial support from the Vilho, Yrj\"o and Kalle V\"ais\"al\"a Foundation (M.K.) and from the Jenny and Antti Wihuri Foundation (H.H.), and the Academy of Finland Project No. 330448 (K.J.E.). This research was funded as a part of the Center of Excellence in Quark Matter of the Academy of Finland (Projects No. 346325 and 364192). This research is part of the European Research Council Project No. ERC-2018-ADG-835105 YoctoLHC, and the European Union's Horizon 2020 research and innovation program under grant agreement No. 824093 (STRONG-2020). We acknowledge the computation resources from the Finnish IT Center for Science (CSC), project jyy2580, and from the Finnish Computing Competence Infrastructure (FCCI), persistent identifier urn:nbn:fi:research-infras-2016072533.

\end{document}